\documentstyle[12pt]{article}
\newcommand{\be}{\begin{equation}}
\newcommand{\ee}{\end{equation}}
\newcommand{\ba}{\begin{eqnarray}}
\newcommand{\ea}{\end{eqnarray}}

\newcommand{\bi}{\bibitem}
\begin{document}
\begin{center}
{\bf\Huge{KILLING-YANO TENSORS }}
\end{center}
\begin{center}
{\bf\Huge{AND NAMBU TENSORS}}
\end{center}
\begin{center}
 Dumitru Baleanu\footnote{after 1st September  1999 at
Institute of Space Sciences,Bucharest-Magurele, Romania}
\footnote{e-mail:baleanu@thsun1.jinr.ru, baleanu@venus.nipne.ro}
\end{center}
\begin{center}
JINR, Bogoliubov Laboratory of Theoretical Physics
\end{center}
\begin{center}
141980 Dubna,Moscow Region, Russia
\end{center}
\begin{abstract}
We present the conditions when a Killing-Yano tensor  becomes a
Nambu tensor.
We have shown that in the flat space case all  Killing-Yano tensors
are  Nambu tensors.
 In the case of Taub-NUT metric and Kerr-Newmann metric we found that a
Killing-Yano tensor of order two generate a Nambu tensor of order
three.
\end{abstract}

\newpage\setcounter{page}2

\section{Introduction}

 Killing tensors are indispensable tools in the quest for exact
 solutions in many branches of general relativity as well as classical
mechanics.
 Killing tensors can also be important for solving the equations of
motion in particular space-time.The notable example is the Kerr-Newmann
metric which admits a second rank Killing tensor [1].
 In [2] third rank Killing tensors in $(1+1)$-dimensional
geometries were investigated and classified.
 When a manifold admits a Killing-Yano
tensor  we can construct a new constant of motion in the case of
geodesic motion [3].
It was a big success of Gibbons at
all. to have been able to shown that the Killing-Yano tensors , which
had long been known for relativistic systems as a rather mysterious
structure, can be understood as an object generating ''non-generic
symmetry'', i.e. supersymmetry appearing only in the specific
space-time [4].
Exact solutions of Einstein's equations typically admits a number of Killing
vector fields.Some of these vector fields may be motivated by physical
consideration.The higher rank Killing tensors are only geodesic symmetries.
They have no obvious geometric interpretation.An interesting examples of
 Einstein's metrics with Killing-Yano tensors are Taub-NUT
metric and Kerr-Newmann metric.

In 1973 Nambu proposed a profound generalization of
classical Hamiltonian mechanics [5].
The fundamental principles of a canonical form of
Nambu's generalized mechanics , similar to the invariant geometrical
form of Hamiltonian mechanics , has been given by Leon Takhtajan [6].
Nambu's generalization of mechanics is based
upon a higher order $(n\ge2)$ algebraic structure defined on a phase
space M.
In [7] was demonstrated that several Hamiltonian systems possessing
dynamical symmetries can be realized in the Nambu formalism of
generalized mechanics.\\
Despite the elegance  of Nambu's
mechanics , it turns out be somewhat restrictive  with many basic
problems waiting to be solved.

 The main aim of this paper is to
investigate the  connection between Killing-Yano tensors  and Nambu
tensors.\\
 The plan of this paper is as follows.\\
In Sec.2 we summarize
the relevant equations for the Killing-Yano tensors and the associated
Killing tensors of order r.  In Sec.3  a brief review about
Nambu's mechanics is presented.  In Sec.4 we investigate the connection
between Killing-Yano tensors and Nambu tensors.  In Sec.5  the results
 in the case of the flat space , Taub-NUT metric and Kerr-Newmann
metric  are presented. Sec.6 contains our comments and concluding
remarks.
\section{Killing-Yano tensors}
  Let us consider a M  a manifold of dimension N with a  metric
$g_{\mu\nu}$ and no torsion.\\
An r-form field $(1\le r\le N)$ $\eta_{\mu_{1}\cdots \mu_{r}}$ is said
to be a Killing-Yano tensor [3] of order r iff
\be\label{yano}
 D_{({\mu_{1}}}\eta_{\mu_{{2})}\cdots \mu_{r+1}}=0,
\ee

where $D_{\mu}$ denotes the covariant derivative and the parenthesis
denote complete symmetrization over the components indices.
According to (\ref{yano}) the $(r-1)$ form field
\be
l_{\mu_{1}\cdots \mu_{r-1}}=\eta_{\mu_{1}\cdots\mu_{{r-1}m}}p^{m},
\ee
is parallel transported along affine parametrized  geodesics with
tangent field  $p^{a}$.
Let us consider $\eta_{\mu_{1}\cdots\mu_{r}}$ a Killing-Yano
tensor of order r , then  a tensor field
\be\label{ab}
K_{ab}=\eta_{a \mu_{2}\cdots
\mu_{r}}{\eta^{\mu_{2}\cdots\mu_{r}}}_{b},
\ee
 is symmetric and proves to
be a Killing tensor called the associated Killing tensor.
We know that a symmetric tensor field $K_{\mu_{1}\cdots\mu_{r}}$
is called a Killing tensor of valence r iff
\be\label{kil}
D_{(\mu_{1}}{K_{\mu_{2}\cdots\mu_{r+1})}}=0,
\ee
where the parenthesis denote complete symmetrization over the components
indices.
The equation (\ref{kil}) ensures that
$K_{\mu_{1}\cdots\mu_{r}}p^{\mu_{1}}\cdots p^{\mu_{r}}$ is a first integral of
the geodesic equation.
{ }For
every Killing-Yano tensor $\eta_{\mu_{1}\cdots\mu_{r}}$ of order r,
in the case of the geodesic motion,
from (\ref{ab},\ref{kil}) we construct  a constant of motion.
Then
the quantities
$H=g_{\mu\nu}p^{\mu}p^{\nu}$,
$K^{l}=\eta_{\mu_{1}\cdots\mu_{r-1}a}{\eta^{\mu_{1}\cdots\mu_{r-1}}}_{b}p^{a}p^{b}$
(for  $l=3,\cdots N$) are conserved quantities in the case of the
geodesic motion.

\section{Nambu Mechanics}
 In this section we would like to make a brief review of the Nambu's
mechanics.
 M is called a Nambu-Poisson manifold [6] if there exists a
R-multilinear map \be \{,\cdots,\}:[C^{(\infty)}]^{\otimes}\to
C^{\infty}(M) \ee called a Nambu bracket of order n such that $\forall
f_{1},f_{2}\cdots,f_{2n-1}$
\be\label{eq2}
\{f_{1},\cdots,f_{n}\}=(-)^{\epsilon(\sigma)}\{f_{\sigma(1)},\cdots,f_{\sigma(n)}\},
\ee
\be\label{eq3}
\{f_{1}f_{2},f_{3},\cdots,f_{n+1}\}=f_{1}\{f_{2},f_{3},\cdots,f_{n+1}\}
+\{f_{1},f_{3},\cdots,f_{n+1}\}f_{2},
\ee
and
\ba\label{fundam}
&\{\{f_{1},\cdots,&f_{n-1},f_{n}\}, f_{n+1},\cdots,f_{2n-1}\}
+\{f_{n},\{f_{1},\cdots f_{n-1},f_{n+1}\},
f_{n+2},\cdots,f_{2n-1}\}\cr
&+\cdots+&\{f_{n},\cdots,f_{2n-2},
\{f_{1},\cdots,f_{n-1},f_{2n-1}\}\}\cr
&=& \{f_{1},\cdots,f_{n-1},\{f_{n},\cdots,f_{2n-1}\}\},
\ea
where $\sigma\in S_{n}$- the symmetric group of n elements and
$\epsilon(\sigma)$ is its parity .Equations (\ref{eq2},\ref{eq3})
are the standard skew-symmetry and derivation properties found for
the ordinary $(n=2)$ Poisson bracket, whereas (\ref{fundam}) is a
generalization of the Jacobi identity and was called [6] the
fundamental identity.
 The dynamics on a Nambu -Hamiltonian manifold M (i.e.
a phase space) is determined by n-1 so called Nambu-Hamiltonians
$H_{1},\cdots H_{{n-1}}\in C^{\infty}(M)$ and is governed by the
following equations of motion [6] \be {df\over
dt}=\{f,H_{1},\cdots,H_{n-1}\},{\forall f}\in C^{\infty}(M) \ee The
Nambu bracket is geometrically realized by the Nambu tensor field
$\eta\in\wedge^{n}TM $, a section of the n-fold exterior power
$\wedge^{n}TM $ of a tangent bundle TM, such that
\be
\{f_{1},\cdots,f_{n}\}=\eta\{df_{1},\cdots,df_{n}\}.
\ee
In local coordinates $(x^{1},\cdots,x^{n})$ it becomes
\be
\eta=\eta^{i_{1}\cdots i_{n}}(x){\partial\over\partial
 x^{i_{1}}}\wedge\cdots\wedge{\partial\over\partial x^{i_{n}}}.
\ee
The fundamental identity (\ref{fundam}) is equivalent to the algebraic
constraint equation for a Nambu tensor ${\eta^{i_{1}\cdots i_{n}}}$
\be\label{alg}
N^{i_{1}\cdots i_{n} j_{1}\cdots j_{n}} +N^{j_{1}
i_{2}i_{3}\cdots i_{n}i_{1}j_{2}j_{3}\cdots j_{n}}=0 ,\ee where \ba
&N^{i_{1}i_{2}\cdots i_{n}j_{1}j_{2}\cdots j_{n}}:&
=\eta^{i_{1}i_{2}\cdots i_{n}}
\eta^{j_{1}j_{2}\cdots j_{n}} + \eta^{j_{n}i_{1}i_{3}\cdots
i_{n}}\eta^{j_{1}j_{2}\cdots j_{n-1}i_{2}}\cr &+\cdots +&
\eta^{j_{n}i_{2}i_{3}\cdots i_{n-1}i_{1}}\eta^{j_{1}j_{2}\cdots
j_{n-1}i_{n}}- \eta^{j_{n}i_{2}i_{3}\cdots i_{n}}\eta^{j_{1}j_{2}\cdots
j_{n-1}i_{1}},
\ea
and one differential constraint equation [see Refs.8,9 for more
details]
\ba\label{dif}
&D^{i_{2}\cdots i_{n}j_{1}\cdots j_{n}}:=& \eta^{k
i_{2}\cdots i_{n}}{\partial _{k}}\eta^{j_{1}j_{2}\cdots j_{n}}
+\eta^{j_{n}k i_{3}\cdots i_{n}}{\partial_{k}}\eta^{j_{1}j_{2}\cdots
j_{n-1}i_{2}}\cr
&+ \cdots +&
\eta^{j_{n}i_{2}i_{3}\cdots i_{n-1}k}{\partial
_{k}}\eta^{j_{n}i_{2}i_{3}\cdots i_{n-1}k}-
\eta^{j_{1}j_{2}\cdots j_{n-1}k}{\partial_{k}}\eta^{j_{n}i_{2}i_{3}\cdots
i_{n}}=0.
\ea

 It has been shown [10]  that  Nambu tensors are decomposable (as conjectured in [9])
which in particular means that they can be written as determinants of
the form (for more details see Ref.11)
\be\label{imp}
\eta^{i_{1}\cdots
i_{n}}=\epsilon_{\alpha_{1}\cdots\alpha_{n}}v^{i_{1}{\alpha_{1}}}\cdots
v^{i_{n}{\alpha_{n}}}. \ee
\section{Nambu and Killing-Yano tensors}
In this section we present the connection between Killing-Yano
tensors and Nambu tensors.\\
Let us suppose that a N-dimensional manifold M with a metric and no
torsion admits a covariant constant  Killing-Yano tensor $\eta_{\mu\nu}$.
We define $\eta^{\mu_{i}\alpha_{j}}$ as\\
$\eta^{\mu_{i}\alpha_{j}}=g^{\mu_{i}\lambda}g^{\alpha_{j}\beta}\eta_{\lambda\beta}$
for $i,j=1,\cdots r$.Then the following two propositions hold.\\

{\bf Proposition 1}\\

  If $\eta_{\mu\nu}$ satisfies $D_{\lambda}\eta_{\mu\nu}=0$ then
 \be\label{propo1}
 \eta^{\mu_{1}\cdots\mu_{r}}=\epsilon_{\alpha_{1}\cdots\alpha_{r}}\eta^{\mu_{1}\alpha_{1}}\cdots\eta^{\mu_{r}\alpha_{r}},
 \ee
 is a covariant constant  tensor.\\

 {\bf Proof.}\\

  Let us denote by $\eta_{\mu\nu}$  a covariant constant
Killing-Yano tensor .
Because $D_{\lambda}\eta_{\mu\nu}=0$ and $D_{\lambda}g_{\mu\nu}=0$
we  conclude from (\ref{yano}) that
\be\label{dcov}
D_{\lambda}\eta^{\mu\nu}=0. \ee
On the other hand $\eta^{\mu_{1}\cdots\mu_{r}}$ from (\ref{propo1}) is
antisymmetric by construction.  Using (\ref{propo1}) and (\ref{dcov}) we
can deduce immediately that $D_{\lambda} \eta^{\mu_{1}\cdots\mu_{r}}=0$.
{\bf q.e.d.}\\

 {\bf Proposition 2}\\

 A  covariant constant  tensor $\eta^{\mu_{1}\cdots\mu_{r}}$
in  a determinant form is a Nambu tensor.\\

{\bf Proof.}\\

 Let us suppose that $\eta^{\mu_{1}\cdots\mu_{r}}$  is a covariant constant
 tensor
in the form (\ref{imp}).
On the other hand we know that any antisymmetric tensor of order r
$\eta^{\mu_{1}\cdots\mu_{r}}$
  automatically satisfies (\ref{alg}) by $N^{i_{1}\cdots
i_{n}j_{1}\cdots j_{n}}=0$ (for more details see Ref.11).\\
 From
 $D_{k}\eta^{\mu_{1}\cdots\mu_{r}}=0$
we get
\be\label{dife}
{\partial\eta^{\mu_{1}\cdots\mu_{r}}\over\partial x^{k}}=
-\Gamma^{\mu_{1}}_{km}\eta^{m\cdots\mu_{r}}-\cdots -
\Gamma^{\mu_{r}}_{km}\eta^{\mu_{1}\cdots m}. \ee
Taking into account (\ref{dife}) the relation (\ref{dif}) becomes
\ba\label{rez}
D^{i_{2}\cdots
i_{n}j_{1}\cdots j_{n}}:&=&\eta^{ki_{2}\cdots
i_{n}}[\Gamma^{j_{1}}_{km}\eta^{m\cdots j_{n}}+
\cdots +\Gamma^{j_{n}}_{km}\eta^{j_{1}\cdots}]+\cr
&\cdots +& \eta^{j_{n}\cdots
i_{n-1}k}[\Gamma^{j_{1}}_{km}\eta^{m\cdots i_{n}}+\cdots +
\Gamma^{i_{n}}_{km}\eta^{j_{1}\cdots m}]\cr
&-&\eta^{j_{1}\cdots j_{n-1}k}[\Gamma^{j_{n}}_{km}\eta^{mi_{2}\cdots
i_{n}}+\cdots +\Gamma^{i_{n}}_{km}\eta^{j_{n}\cdots m}].
\ea
Here $\Gamma^{i_{1}}_{km}$ denotes a Christoffel symbol.
 Because  $N^{i_{1}\cdots
i_{n}j_{1}\cdots j_{n}}=0$ after some calculations we found that
(\ref{rez}) becomes
\be\label{fin} D^{i_{2}\cdots i_{n}j_{1}\cdots
j_{n}}:= \Gamma^{j_{1}}_{km}\eta^{j_{n}i_{2}\cdots
i_{n}}\eta^{mj_{2}\cdots j_{n-1}k} +\cdots +
\Gamma^{j_{n-1}}_{km}\eta^{j_{n}i_{2}\cdots i_{n}}\eta^{j_{1}\cdots
mk}.  \ee Using the fact that the manifold M has  no torsion we
conclude that (\ref{fin}) is identically zero.  Then
$\eta^{\mu_{1}\cdots\mu_{r}}$ becomes a Nambu tensor.\\
{\bf q.e.d.}

An interesting case arises when a manifold M
with a metric $g_{\mu\nu}$ and  torsion free admits a non-trivial
Killing-Yano tensor $\eta_{\mu\nu}$.
And in this case we
 can construct a tensor  in
a determinant form  as
$\eta^{\mu_{1}\cdots\mu_{r}}=\epsilon_{\alpha_{1}\cdots\alpha_{r}}\eta^{\mu_{1}\alpha_{1}}\cdots\eta^{\mu_{r}\alpha_{r}}$
	 ,where

$\eta^{\mu_{i}\alpha_{j}}=g^{\mu_{i}\lambda}g^{\alpha_{j}\beta}\eta_{\lambda\beta}$.
If this tensor
satisfies (\ref{dif}) we said that a Killing-Yano tensor of order two
generates a Nambu tensor of order r.

\section{Examples}
 In this
section we present the connection between Killing-Yano tensors and
Nambu tensors in the cases of flat space, Taub-NUT metric and
Kerr-Newmann metric.
\subsection{Flat Space} Let $E^{n+1}$ be an
Euclidean space and ${x^{\lambda}}$ $(\lambda=1\cdots n+1)$ an
orthogonal coordinate system.  A Killing-Yano tensor of order r is an
antisymmetric  tensor such that \be\label{flat}
\partial_{(\mu}\eta_{\nu_{1})\nu_{2}\cdots\nu_{r}}=0.
\ee
Here $r=3,\cdots n+1 $ and $\eta_{\nu_{1}\cdots\nu_{n+1}}$ is
proportional with $\epsilon_{\mu_{1}\cdots\mu_{n+1}}$.
 The general solution of (\ref{flat}) is
\be\label{ten}
\eta_{\nu_{1}\cdots\nu_{r}}=x^{\nu}g_{\nu\nu_{1}\cdots\nu_{r}}
+h_{\nu_{1}\cdots\nu_{r}}.
\ee
where $g_{\nu\nu_{1}\cdots\nu_{r}}$ and $h_{\nu_{1}\cdots\nu_{r}}$ are constant
antisymmetric tensors.
Because euclidean space has  dimension n+1
we get n-1 Killing-Yano tensors of the form (\ref{ten}).

 The associate Killing tensor $K_{ab}$  has the following
expression
\ba\label{kila}
&K_{ab}&=\eta_{a\nu_{2}\cdots\nu_{r}}{\eta^{\nu_{2}\cdots\nu_{r}}}_{b}
=x^{\alpha}x^{\beta}(g_{a \alpha\nu_{2}\cdots\nu_{r}}{g^{\nu_{2}\cdots\nu_{r}}}_{b\beta}+
g_{b \alpha\nu_{2}\cdots\nu_{r}}{g^{\nu_{2}\cdots\nu_{r}}}_{a\beta})\cr
&+& x^{\alpha}(g_{a \alpha\nu_{2}\cdots\nu_{r}}{h^{\nu_{2}\cdots\nu_{r}}}_{b} +
+ g_{b \alpha\nu_{2}\cdots\nu_{r}}{h^{\nu_{2}\cdots\nu_{r}}}_{a} +
 {g^{\nu_{2}\cdots\nu_{r}}}_{\alpha b}h_{a \nu_{2}\cdots\nu_{r}} +
 {g^{\nu_{2}\cdots\nu_{r}}}_{\alpha a}h_{b \nu_{2}\cdots\nu_{r}})\cr
&+& h_{a\nu_{2}\cdots\nu_{r}}{h^{\nu_{2}\cdots\nu_{r}}}_{b}
+ h_{b\nu_{2}\cdots\nu_{r}}{h^{\nu_{2}\cdots\nu_{r}}}_{a}.
\ea
From (\ref{ten}) and (\ref{kila}) we get
  n-1 constants of
motions  of the form $ K=K_{ab}p^{a}p^{b}$.

 If we take into account  the Hamiltonian H we have n
constants of motion for the flat space case.
\subsection{Taub-NUT metric}
The four-dimensional Taub-NUT metric
depends on a parameter $m$ which can be positive or negative, depending
on the application; for $m>0$ it represents a nonsingular solution of
the self-dual Euclidean equation and as such is interpreted as a
gravitational instanton. The standard form of the line element is

\begin{eqnarray}\label{ade}
ds^2 &=& \left(1+\frac{2m}{r}\right)(dr^2+r^2d\theta^2+r^2\sin^2\theta d\varphi^2)\nonumber\\ &&
+\frac{4m^2}{1+2m/r}(d\psi + \cos\theta d\varphi)^2.
\end{eqnarray}
The symmetries of extended Taub-NUT metric and dual metrics were investigated in [13].
The metric (\ref{ade}) admits four Killing-Yano tensors [12].
Three of these,
denoted by $f_i$ are special because they are covariant constant. In the
two-form notation the explicit expressions are [12]
\begin{equation}\label{kyf}
f_i=4m(d\psi +\cos \theta d\varphi )dx_i-\epsilon _{ijk}(1+{\frac{2m}r}%
)dx_j\wedge dx_k,
\end{equation}
where the $dx_i$ are standard expressions in terms of the 3-dimensional
spherical co-ordinates $(r,\theta ,\varphi )$.
The fourth Killing-Yano tensor has the following form
\begin{equation}\label{ky}
Y=4m(d\psi
+\cos \theta d\varphi )\wedge dr+4r(r+m)(1+{\frac r{2m}})\sin \theta
d\theta \wedge d\varphi.
\end{equation}
Because $D_{\lambda}f_{i}=0$ from  Proposition 1  we get a
Killing-Yano tensor of order three and using  Proposition 2 we
found that it is a Nambu tensor.We have the following results:

{\bf Case 1.}\\

For i=1 we found three
independent components of Nambu tensor $\eta^{i_{1}i_{2}i_{3}}$
\ba
&\eta^{r\theta\psi}&=4{\sin\varphi\over mr\sin\theta{(r+2m)^{2}}},
\eta^{r\varphi\psi}=-4{\cos\varphi\over
mr\sin^{2}\theta{(r+2m)^{2}}},\cr
&\eta^{\theta\varphi\psi}&=-4{\sin\varphi\over
mr^{2}\sin\theta{(r+2m)^{2}}}.
\ea

{\bf Case 2.}\\

When i=2  we have obtained three independent components of Nambu tensor
\ba
&\eta^{r\theta\psi}&=-4{\cos\varphi\over mr\sin\theta{(r+2m)^{2}}},
\eta^{r\varphi\psi}=4{\sin\varphi\over mr\sin^{2}\theta{(r+2m)^{2}}},\cr
&\eta^{\theta\varphi\psi}&=-4{\sin\varphi\over
mr^{2}\sin\theta{(r+2m)^{2}}}. \ea

{\bf Case 3.}\\

For i=3  we have only two independent components for Nambu tensor

\be
\eta^{r\varphi\psi}=-4{\cos\theta\over mr\sin^{2}\theta{(r+2m)^{2}}},
\eta^{\theta\varphi\psi}=-4{\cos\theta\over
mr^{2}\sin^{2}\theta{(r+2m)^{2}}}. \ee

 The Killing-Yano tensor Y from (\ref{ky}) has not covariant derivative
zero but we can construct and in this case a  tensor
$\eta^{\mu_{1}\mu_{2}\mu_{3}}$ in a determinant form which satisfies
(\ref{dif}).
Then  the Killing-Yano tensor Y generates a Nambu tensor of order three.
By direct calculations we found that in this case the component of  Nambu
tensor is \be
\eta^{\theta\varphi\psi}={{2(r^{4}+4r^{3}m +6r^{2}m^{2}+4rm^{3}
+m^{4})}\over { m^{3}r^{2}[-\sin^{2}{\theta}(r^{4} +16m^{4})
+8mr\cos^{2}\theta(r^{2} +3rm +4m^{2})]}}. \ee
\subsection{Kerr-Newmann
metric}

The Kerr-Newmann geometry describes a charged spinning black hole; in a
standard choice of coordinates the metric is given by the following line
element [1]:
\begin{eqnarray}
 ds^2=-\frac{\Delta}{\rho^2}\left[dt-a\sin^2(\theta)d\varphi\right]^2+
\frac{\sin(\theta)^2}{\rho^2}\left[(r^2+a^2)d\varphi -adt\right]^2
+\frac{\rho^2}{\Delta}dr^2+\rho^2d\theta^2.
\end{eqnarray}
       Here
\begin{eqnarray}
\Delta=r^2+a^2-2Mr+Q^2,
\rho^2=r^2+a^2\cos^2\theta,
\end{eqnarray}
with $Q$ the background electric charge, and $J=Ma$ the
total angular momentum. The expression for $ds^2$ only describes the fields
{\em outside} the horizon, which is located at
\begin{eqnarray}
r=M+(M^2-Q^2-a^2)^{1/2}.
\end{eqnarray}
The Killing -Yano tensor for the Kerr-Newmann is defined by [1]
\begin{eqnarray}\label{ker}
\frac{1}{2}f_{\mu\nu}dx^{\mu}\wedge dx^{\nu}=\nonumber
a\,\cos\theta\, dr\wedge(dt-a\,\sin^2\theta\, d\phi)\nonumber\\
+r\,\sin\theta\, d\theta\wedge[-a\,dt+(r^2+a^2)\,d\phi].
\end{eqnarray}
The Killing-Yano tensor (\ref{ker}) has not covariant derivative zero
but  generates a Nambu tensor of order three.
 We found in this case the following non-vanishing components
\be
\eta^{r\varphi t}={{\cos^{6}\theta}\sin\theta
a^{7}r^{3}\over\sin^{2}\theta(r^{2}+a^{2})^{2}},
\eta^{\theta\varphi t}={{r^{6}\cos^{3}\theta a^{3}}\over
\sin^{2}\theta(r^{2}+a^{2})^{2}}. \ee

\section{Concluding remarks}
 The physical
and geometrical interpretation  of Killing-Yano tensors  of rang higher
than two is not clarified [14].
The existence of a Killing-Yano tensor  is both
a necessary and sufficient condition for the existence of a new
supersymmetry of the spinning space [4,15].

In this paper we investigate the connection between Killing-Yano
tensors and Nambu tensors.
If the manifold admits a covariant constant Killing-Yano
tensor of order two  we have constructed a covariant constant
 tensor of higher order.
We have proved that a covariant constant  tensor,
in a determinant form, on a given N-dimensional manifold
with  metric and no torsion, is a Nambu tensor.
The flat space case  was investigated in details.  In
this case  the  general solution of Killing-Yano equations was presented
and we have proved that all Killing-Yano tensors are Nambu tensors.
The  expression for the associated Killing tensors was obtained.
Taub-NUT metric admits four Killing-Yano tensors of order two .
Three of these,
denoted by $f_i$ are covariant constant.Using  Proposition 1
and  Proposition 2 , for each Killing-Yano tensor $f_{i}$, we found
the corresponding Nambu tensor of order three.
The fourth Killing-Yano tensor Y is not covariant constant. In this
case we have constructed a tensor of order three in a determinant form
.We found that this tensor is a Nambu tensor because it
satisfies  (\ref{dif}).  Kerr-Newmann metric admits one non-trivial
 Killing-Yano tensor of order two.  And in this case we have
constructed a tensor of rank three in a determinant form and we have
proved that it is a Nambu tensor.  Then all Killing-Yano tensors of
order two in the case of Taub-NUT and Kerr-Newmann metric generate
Nambu tensors of order three.

\section{Acknowledgments}
I am grateful to M. Flato  and D.Alekseevsky for useful discussions.
I would like to thank TUBITAK for financial
support  and METU for the hospitality during his
working stage at Department of Physics.

\end{document}